# Feasibility of Extracting Skin Nerve Activity from Electrocardiogram Recorded at A Low Sampling Frequency*

Youngsun Kong, *Member, IEEE*, Farnoush Baghestani, *Student Member, IEEE*, I-Ping Chen, and Ki Chon, *Fellow, IEEE*

*Abstract*—Skin nerve activity (SKNA) derived from electrocardiogram (ECG) signals has been a promising non-invasive surrogate for accurate and effective assessment of the sympathetic nervous system (SNS). Typically, SKNA extraction requires a higher sampling frequency than the typical ECG recording requirement ($\geq 2$ kHz) because analysis tools extract SKNA from the 0.5–1 kHz frequency band. However, ECG recording systems commonly provide a sampling frequency of 1 kHz or lower, particularly for wearable devices. Our recent power spectral analysis exhibited that 150–500 Hz frequency bands are dominant during sympathetic stimulation. Therefore, we hypothesize that SKNA can be extracted from ECG sampled at a lower sampling frequency. We collected ECG signals from 16 participants during SNS stimulation and resampled the signals at 0.5, 1, and 4 kHz. Our statistical analyses of significance, classification performance, and reliability indicate no significant difference between SKNA indices derived from ECG signals sampled at 0.5, 1, and 4 kHz. Our findings indicate that conventional ECG devices, which are limited to low sampling rates due to resource constraints or outdated guidelines, can be used to reliably collect SKNA if muscle artifact contamination is minimal.

*Clinical Relevance*—Our study provides a crucial groundwork for wearable SKNA research in affective and cardiovascular research, which require reliable assessment of SNS.

## I. Introduction

The sympathetic nervous system (SNS) is a branch of the autonomic nervous system and plays a key role in modulating neural activity in response to "fight or flight" conditions, influencing bodily functions, such as increased heart rate, blood pressure, sweating, and others. The SNS has been linked to emotion and stress [1], [2], as well as kidney dysfunction, hypertension, coronary artery disease, and heart failure [3], [4], [5]. Consequently, researchers have used non-invasive markers, such as heart rate variability (HRV) and electrodermal activity (EDA), to objectively assess the affective and cardiovascular conditions [6], [7], [8].

Recent studies have shown that skin nerve activity (SKNA) can be extracted from electrocardiogram (ECG) signals recorded via conventional ECG recording system. SKNA is commonly extracted using a technique called neuECG, which was developed by Kusayama *et al.*, the pioneers of SKNA studies [9]. The neuECG technique involves a bandpass filter with cutoff frequencies of 0.5–1 kHz, followed by rectification and smoothing to derive integrated SKNA (iSKNA). Their study exhibited that iSKNA has comparable dynamics to sympathetic nerve activity recorded from microneurography [9]. These frequency bands, which are higher than those of PQRST waveforms, are thought to originate from the stellate ganglion, a key structure in the SNS [10], [11]. SKNA has been found to increase during cold pressor testing, cognitive stress, pain, as well as better classification performance compared to EDA and HRV [12], [13]. Therefore, SKNA has emerged as a promising non-invasive surrogate for the SNS assessment.

Despite its great potential, it requires a minimum sampling frequency of 2 kHz according to the Nyquist theorem, because the neuECG technique extracts SKNA between 0.5–1 kHz frequency band. This frequency band limits the analysis of ECG signals recorded with a sampling frequency lower than 2 kHz. This may be an issue in which ECG recording systems commonly provide sampling frequency of 1 kHz or lower. The American Heart Association recommends 500 Hz as the minimum sampling frequency [14]. Many ECG recording devices have been designed according to the guideline, hence, sample ECG signals at 1 kHz or lower [15], [16].

The neuECG technique extracts SKNA from the 0.5–1 kHz frequency band because: 1) the predominant frequency band of muscle movement is below 400 Hz and 2) the dynamics of microneurography are often highpass filtered at 700 Hz [9]. However, a recent study showed that muscle noise interferences in SKNA occupy a 95% energy band between 508 – 898 Hz [17]. In addition, our recent power spectral analysis also exhibited that dominant frequency band activated during SNS stimulation in ECG-derived SKNA is between 150 and 500 Hz [18]. To summarize, the lower frequencies (150–500 Hz) are more indicative of SNS activity, while muscle noise is an issue not only below 400 Hz but also in the 0.5–1 kHz frequency band. Therefore, we hypothesized that SKNA can be extracted from ECG signals recorded at sampling frequencies of 1 kHz or lower, particularly when collected without the muscle artifact contamination. To investigate this, we extracted SKNA indices from ECG signals resampled at 0.5, 1, and 4 kHz during SNS stimulation.

## II. Methods

### A. Data collection

Eight males and eight females (20–57 y/o) were recruited to perform three tasks which were designed to invoke SNS activity, including the Valsalva maneuver (VM), Stroop test

*Research was supported by the National Institute of Dental & Craniofacial Research of the National Institutes of Health under Award Number F32DE033566.

Y. K, F. B., and K. C. are with the department of biomedical engineering at the University of Connecticut, Storrs, CT 06269 USA (corresponding author to provide phone: 860-486-6975; e-mail: yskong224@gmail.com).

I. C. is with the department of Oral Health and Diagnostic Sciences, University of Connecticut Health, Farmington, CT 06030, USA

(ST), and a thermal grill (TG) test. VM was performed by a deep breath followed by forcefully exhaling against a closed airway while closing the mouth for 30 seconds, three to four times for each subject. During ST, a series of screens consisting of the following six words on a smartphone tablet are shown at 1-3 second intervals: "Red", "Blue", "Green", "Yellow", "Purple", or "Black". The text color and the screen background color were presented randomly, which could be different from the six words. Participants subvocalized the words' color. Finally, TG involved two grills, where each grill consists of interlaced copper tubes with warm (40–50 °C) and cool water (~18 °C), which can safely induce various intensities of heat pain sensation due to the temperature contrasts. The two grills were adjusted by changing the warm water temperature to induce pain levels 4 – 6 and pain levels > 7 out of the 10-point visual analog scale (VAS), respectively. Participants placed their left hands on the grills while blindfolded, and the location of the grills was adjusted using a wheeled table. Each participant underwent a randomized sequence of 6 stimuli with 3 stimuli for each thermal grill, with an interstimulus interval of approximately 40 seconds. Participants reported pain levels on a 0–10 VAS scale.

Participants were asked to refrain from any stimulants starting 24 hours prior to the start of the experiment. ECG signals were collected at 10 kHz from the following electrode locations using BioAmp with a PowerLab device (ADInstrument, Sydney, Australia): 1) both inner wrists (Lead 1) and 2) the upper left side of the chest and below the right rib cage (Lead III), which is referred to as Channel 1 and Channel 2 for the rest of the paper. This research complied with tenets of the Declaration of Helsinki and was approved by the Institutional Review Board at the University of Connecticut.

### B. Preprocessing

Two time-series SKNA signals were calculated: integrated SKNA (iSKNA) and time-varying index of SKNA (TVSKNA). TVSKNA has shown higher sensitivity and reliability in assessing SNS activity compared to iSKNA [18]. First, ECG signals were resampled at three sampling frequencies: 4, 1, and 0.5 kHz. Then, a bandpass filter was applied for iSKNA computation with cutoff frequencies as noted in Table I. For TVSKNA computation, a highpass filter was applied at a cutoff frequency at 150 Hz. Additionally, a series of notch filters was applied to remove noise frequencies identified by power spectral analysis and our visual inspection, which were likely due to artifact contamination (e.g., equipment noise or environmental interference).

TABLE I. COMPARED SKNA SAMPLING FREQUENCIES

| Resampled at | Band frequency interest (iSKNA) | (TVSKNA) | Remarks |
|---|---|---|---|
| 4 kHz [18] | 500–1000 Hz | 480–1,120 Hz | Reference |
| 1 kHz | 250–500 Hz | 240–480 Hz | |
| 0.5 kHz | 150–250 Hz | 160–240 Hz | |

### C. iSKNA and TVSKNA computation

After bandpass filtering, iSKNA was derived by rectifying the signals, followed by a moving average filter with a 100 ms window (Figure 1a and Figure 2).

The TVSKNA computation consists of three steps (Figure 1b and Figure 2): 1) signal decomposition using variable frequency complex demodulation to obtain time-frequency spectrum (TFS) and reconstruction by summing the TFS based on Table II [19], 2) estimation of instantaneous amplitude using the Hilbert transform, and 3) smoothing using a moving average filter with a 100 ms window.

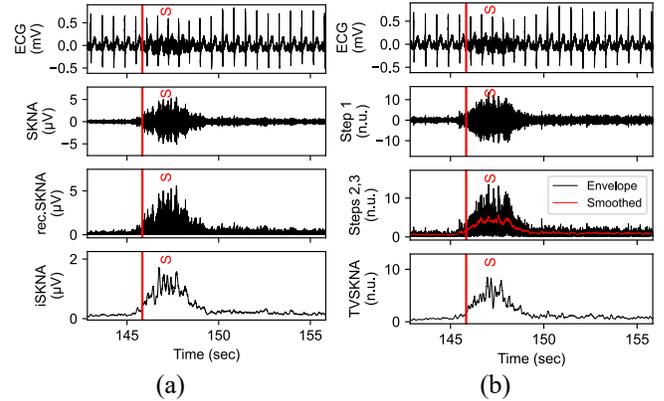

Figure 1. Example of iSKNA and TVSKNA computation from ECG resampled at 4 kHz (VM).

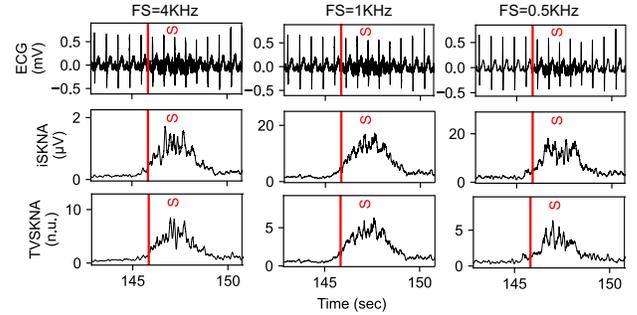

Figure 2. Example of iSKNA and TVSKNA derived from ECG signals resampled at 4, 1, and 0.5 kHz (VM).

TABLE II. CENTER FREQUENCY OF VFCDM COMPONENTS (HZ)

| Decomposed Components | FS = 4 kHz | FS = 1 kHz | FS = 0.5 kHz |
|---|---|---|---|
| 1 | 80 (0–160) | 20 (0–40) | 10 (0–20) |
| 2 | **240 (160–320)** | 60 (40–80) | 30 (20–40) |
| 3 | **400 (320–480)** | 100 (80–120) | 50 (40–60) |
| 4 | **560 (480–640)** | 140 (120–160) | 70 (60–80) |
| 5 | **720 (640–800)** | 180 (160–200) | 90 (80–100) |
| 6 | **880 (800–960)** | 220 (200–240) | 110 (100–120) |
| 7 | **960 (1040–1120)** | 260 (240–280) | 130 (120–140) |
| 8 | 1200 (1120–1280) | **300 (280–320)** | 150 (140–160) |
| 9 | 1360 (1280–1440) | **340 (320–360)** | 170 (160–180) |
| 10 | 1520 (1440–1600) | **380 (360–400)** | 190 (180–200) |
| 11 | 1680 (1600–1760) | **420 (400–440)** | 210 (200–220) |
| 12 | 1840 (1760–1920) | **460 (440–480)** | 230 (220–240) |

Bold fonts indicate the decomposed components used to reconstruct.

### E. Performance Metrics and Statistics

Both iSKNA and TVSKNA indices were calculated from baseline and the SNS task segments. Segment sizes were 30, 120, and 10 seconds for VM, ST, and the TG experiment, respectively. For TG, segments were categorized into clinically non-significant pain (CSP-, $0 < VAS < 4$) and clinically significant pain (CSP+, $VAS \geq 4$). From each segment, maximum, mean, and standard deviation of iSKNA (maxSKNA, aSKNA, and vSKNA, respectively), and those of TVSKNA were computed.

Each index was evaluated for statistical significance, classification performance, and reliability. To determine if the differences between baseline and the SNS task are statistically significant, linear mixed-effects (LMM) models were fitted.

P-values < 0.05 were considered to be statistically significant. Then, Cohen's *d* values were calculated using the fixed effect estimate for the group variable from the LMM models. Also, the area under the curve (AUC) values from the receiver operating characteristic curves were calculated to evaluate the classification performance. Finally, the intraclass correlation coefficients (ICC) were computed to assess variability and reliability of each index across participants.

## III. RESULTS

We excluded some segments due to poor data quality, including sensor and experimenter errors (Table III).

TABLE III.  SEGMENT NUMBERS (AVAILABLE / TOTAL)

|  |  | VM | ST | TG |
|---|---|---|---|---|
| Channel 1 | Participants | 15/16 | 14/16 | 15/16 |
|  | Segments | 47/52 | 14/16 | 135/144 |
| Channel 2 | Participants | 15/16 | 14/16 | 15/16 |
|  | Segments | 49/52 | 14/16 | 115/144 |

Table IV shows the statistical analysis of the iSKNA indices. In terms of significance tests, there was no difference between the sampling frequencies, except for ST Ch. 1. In other words, all indices, except for ST Ch. 1, were consistently non-significant (ST Ch. 2) or significant (the other SNS tasks) between baseline and SNS stimulation across any sampling frequency. Interestingly, lower sampling frequencies showed significant differences in the maxSKNA value of ST Ch. 1 ($p<.05$), while the higher sampling frequency of 4 kHz did not.

While most indices showed large effect sizes ($\geq 0.8$) of Cohen's *d* except for the Stroop test, the 4 kHz sampling frequency generally resulted in lower Cohen's d values compared to both the 1 and 0.5 kHz sampling frequencies. Interestingly, the maxSKNA and aSKNA of ST Ch. 1 showed large effect sizes at the 1 and 0.5 kHz sampling frequencies, but not at the 4 kHz sampling frequency. Regarding the classification performance, AUC values were generally excellent and outstanding for all iSKNA indices ($\geq 0.8$), except for ST [20], AUC values obtained at the 0.5 kHz sampling frequency were slightly smaller compared to other sampling frequencies. For ST Ch. 1, while the 4 kHz sampling frequency showed an acceptable AUC (0.7–0.8), the other sampling frequencies showed AUC values lower than 0.7 with vSKNA and/or maxSKNA. In terms of reliability across participants, most iSKNA indices except for ST showed excellent reliability ($\geq 0.9$) [21]. For ST Ch. 1, aSKNA from all sampling frequencies showed good ICC values (0.75–0.9), while only maxSKNA with the sampling frequency at 1 kHz showed a good ICC value. All indices of ST Ch. 2 exhibited poor ICC values (<0.50).

TVSKNA indices generally showed higher AUC and ICC than those of iSKNA, while the differences between the sampling frequencies were similar to the results of iSKNA (Table V). Cohen's *d* values were generally higher with 1 kHz sampling frequency, followed by those with 0.5 kHz sampling frequency. AUC values were generally excellent and outstanding for all TVSKNA indices ($\geq 0.8$) except for ST. For ST Ch. 1, AUC values of mean SKNA indices at all sampling frequencies appeared to be outstanding ($\geq 0.9$), while the other indices were mostly acceptable (0.7–0.8). In terms of reliability across participants, most TVSKNA indices except for ST showed excellent reliability at all sampling frequencies ($\geq 0.9$). For ST Ch. 1, both max and mean TVSKNA indices showed good or excellent ICC values ($\geq 0.75$) at all sampling frequencies. For standard deviation values, only 0.5 kHz showed an acceptable ICC value, while showing good reliability ($\geq 0.75$).

## IV. DISCUSSIONS AND CONCLUSION

In this study, we tested the feasibility of extracting SKNA from a low sampling frequency. To achieve this, iSKNA and TVSKNA indices were calculated according to our previous power spectral analysis which found that dominant frequency components of SKNA exist between 150 and 500 Hz [18]. Our results indicate no noticeable differences in terms of significance, classification performance, and reliability between sampling frequencies. In some cases, the performance was better with sampling frequencies lower than 4 kHz. This suggests that SKNA can be extracted using ECG recording devices that can sample at 1 kHz, or lower, particularly wearable devices that require lower sampling frequencies due to the resource-constrained environments (e.g., battery, processing power, etc.).

TABLE IV.  STATISTICAL ANALYSIS OF INTEGERATED SKNA (iSKNA) INDICES

|  |  | VM | | | ST | | | CSP- | | | CSP+ | | |
|---|---|---|---|---|---|---|---|---|---|---|---|---|---|
| Channel 1 |  | Cohen's d | AUC | ICC | Cohen's d | AUC | ICC | Cohen's d | AUC | ICC | Cohen's d | AUC | ICC |
| 4 kHz | Max | 2.68** | 1 | 0.99 | 0.5 | 0.7 | 0.67 | 2.2** | 1 | 0.99 | 1.39** | 1 | 0.98 |
| (0.5-1 kHz) | Mean | 1.7** | 0.9 | 0.96 | 0.72* | 0.72 | 0.8 | 2.22** | 0.99 | 0.99 | 2.32** | 1 | 0.98 |
|  | S.D. | 2.51** | 1 | 0.99 | 0.33 | 0.7 | 0.32 | 1.7** | 0.99 | 0.98 | 2.15** | 1 | 0.99 |
| 1 kHz | Max | 4.06** | 1 | 0.99 | 1.12* | 0.77 | 0.87 | 3.2** | 0.98 | 0.99 | 3.07** | 1 | 0.99 |
| (250-500 Hz) | Mean | 2.34** | 0.84 | 0.93 | 1.44* | 0.72 | 0.82 | 3.32** | 0.97 | 0.99 | 4.1** | 0.99 | 0.99 |
|  | S.D. | 4.12** | 1 | 0.99 | 0.39 | 0.66 | 0.05 | 2.39** | 1 | 0.98 | 3.42** | 1 | 0.99 |
| 0.5 kHz | Max | 3.05** | 0.98 | 0.98 | 0.95* | 0.69 | 0.73 | 3.84** | 1 | 0.99 | 2.96** | 1 | 0.99 |
| (150-250 Hz) | Mean | 1.88** | 0.76 | 0.87 | 1.54* | 0.7 | 0.79 | 3.34** | 0.99 | 0.99 | 4.2** | 0.99 | 0.99 |
|  | S.D. | 3.05** | 0.99 | 0.98 | 0.5 | 0.62 | 0.28 | 2.62** | 1 | 0.98 | 3.27** | 1 | 0.99 |
| Channel 2 |  | Cohen's d | AUC | ICC | Cohen's d | AUC | ICC | Cohen's d | AUC | ICC | Cohen's d | AUC | ICC |
| 4 kHz | Max | 2.1** | 0.99 | 0.98 | 0.4 | 0.63 | 0.48 | 1.09* | 0.76 | 0.9 | 0.72* | 0.91 | 0.92 |
| (0.5-1 kHz) | Mean | 1.76** | 0.88 | 0.92 | 0.32 | 0.6 | 0.25 | 1.7** | 0.8 | 0.93 | 1.49** | 0.88 | 0.94 |
|  | S.D. | 2.33** | 1 | 0.98 | 0.2 | 0.6 | 0 | 0.89* | 0.83 | 0.92 | 0.86** | 0.93 | 0.94 |
| 1 kHz | Max | 3.09** | 0.99 | 0.98 | 0.45 | 0.61 | 0.28 | 1.48* | 0.89 | 0.92 | 1.18** | 0.92 | 0.93 |
| (250-500 Hz) | Mean | 3.05** | 0.92 | 0.96 | 0.46 | 0.56 | 0.27 | 2.39** | 0.84 | 0.94 | 2.24** | 0.87 | 0.93 |
|  | S.D. | 3.69** | 0.99 | 0.98 | 0.17 | 0.56 | 0 | 1.08* | 0.88 | 0.89 | 1.36** | 0.93 | 0.94 |
| 0.5 kHz | Max | 2.28** | 0.97 | 0.97 | 0.53 | 0.61 | 0.47 | 2.23* | 0.85 | 0.94 | 0.85* | 0.81 | 0.91 |
| (150-250 Hz) | Mean | 2.76** | 0.89 | 0.95 | 0.46 | 0.58 | 0.27 | 3.04** | 0.83 | 0.94 | 2.55** | 0.83 | 0.92 |
|  | S.D. | 2.69** | 0.97 | 0.97 | 0.11 | 0.46 | 0 | 1.35* | 0.75 | 0.9 | 0.99* | 0.78 | 0.9 |

\* *p*<.05, \*\* *p*<.001. VM: Valsalva maneuver, ST: Stroop test, TG: thermal grill, Max: maxSKNA, Mean: aSKNA, S.D.: vSKNA

TABLE V. STATISTICAL ANALYSIS OF TVSKNA INDICES

| | | VM | | | ST | | | CSP- | | | CSP+ | | |
|---|---|---|---|---|---|---|---|---|---|---|---|---|---|
| Channel 1 | | *Cohen's d* | *AUC* | *ICC* | *Cohen's d* | *AUC* | *ICC* | *Cohen's d* | *AUC* | *ICC* | *Cohen's d* | *AUC* | *ICC* |
| 4 kHz (480-1,120 Hz) | Max | 1.93** | 0.96 | 0.98 | 0.64* | 0.8 | 0.8 | 3.7** | 1 | 1 | 1.4** | 1 | 0.99 |
| | Mean | 1.69** | 0.95 | 0.95 | 0.76* | 0.9 | 0.92 | 1.84** | 1 | 0.99 | 1.95** | 1 | 1 |
| | S.D. | 2.23** | 0.96 | 0.98 | 0.35 | 0.76 | 0.53 | 3.63** | 1 | 1 | 1.76** | 1 | 0.99 |
| 1 kHz (240-480 Hz) | Max | 3.24** | 0.96 | 0.98 | 0.98* | 0.79 | 0.85 | 4.1** | 1 | 1 | 2.87** | 1 | 1 |
| | Mean | 2.17** | 0.93 | 0.94 | 1.7** | 0.95 | 0.95 | 4.83** | 1 | 1 | 4.97** | 1 | 1 |
| | S.D. | 3.47** | 0.96 | 0.98 | 0.41 | 0.69 | 0.17 | 2.67** | 1 | 0.99 | 3.51** | 1 | 0.99 |
| 0.5 kHz (160-240 Hz) | Max | 2.75** | 0.96 | 0.98 | 0.78 | 0.71 | 0.77 | 4.4** | 1 | 0.99 | 2.97** | 1 | 1 |
| | Mean | 1.64** | 0.84 | 0.92 | 1.57** | 0.92 | 0.94 | 4.05** | 1 | 0.99 | 4.79** | 1 | 0.99 |
| | S.D. | 2.97** | 0.96 | 0.98 | 0.84* | 0.78 | 0.8 | 3.37** | 1 | 0.99 | 3.45** | 1 | 0.99 |
| Channel 2 | | *Cohen's d* | *AUC* | *ICC* | *Cohen's d* | *AUC* | *ICC* | *Cohen's d* | *AUC* | *ICC* | *Cohen's d* | *AUC* | *ICC* |
| 4 kHz (480-1,120 Hz) | Max | 2.64** | 1 | 1 | 0.29 | 0.55 | 0.16 | 1.09** | 0.93 | 0.95 | 0.96** | 0.93 | 0.98 |
| | Mean | 1.71** | 0.99 | 0.99 | 0.32 | 0.63 | 0.61 | 1.11** | 0.97 | 0.96 | 1.2** | 0.95 | 0.99 |
| | S.D. | 2.73** | 1 | 0.99 | 0.1 | 0.5 | 0 | 1.45** | 0.97 | 0.97 | 0.97** | 0.92 | 0.98 |
| 1 kHz (240-480 Hz) | Max | 3.32** | 1 | 0.99 | 0.46 | 0.61 | 0.25 | 2.03** | 1 | 0.97 | 1.19** | 0.96 | 0.95 |
| | Mean | 3** | 0.99 | 0.99 | 0.69 | 0.66 | 0.7 | 3** | 0.99 | 0.99 | 2.67** | 0.98 | 0.98 |
| | S.D. | 4.26** | 1 | 0.99 | 0.09 | 0.52 | 0 | 1.4* | 0.88 | 0.93 | 1.39** | 0.92 | 0.95 |
| 0.5 kHz (160-240 Hz) | Max | 2.72** | 1 | 0.99 | 0.66 | 0.7 | 0.65 | 3.65** | 0.99 | 0.98 | 0.83* | 0.96 | 0.93 |
| | Mean | 2.66** | 0.97 | 0.98 | 0.44 | 0.58 | 0.27 | 3.42** | 0.98 | 0.98 | 2.68** | 0.96 | 0.98 |
| | S.D. | 3.2** | 1 | 0.99 | 0.25 | 0.57 | 0 | 2.1** | 0.93 | 0.95 | 1.1** | 0.92 | 0.95 |

* $p<.05$, ** $p<.001$. VM: Valsalva maneuver, ST: Stroop test, TG: thermal grill

Despite this promising results with low sampling frequencies, it is always recommended sampling frequencies at 2 or 3 times the theoretical minimum, because the Nyquist theorem is valid for an infinite sampling interval [14]. Furthermore, our datasets were collected in well-controlled lab environment. In real-world data involving muscle noise, upper frequency bands ($\geq 500$ Hz) may be more appropriate, which is currently unknown. Despite these limitations, which should be addressed in future studies, we suggest that SKNA can be extracted from ECG signals sampled at 0.5 and 1 kHz if muscle artifact contamination is minimal.

REFERENCES

[1] R. W. Levenson, "The Autonomic Nervous System and Emotion," *Emotion Review*, vol. 6, no. 2, pp. 100–112, Apr. 2014, doi: 10.1177/1754073913512003.
[2] J. A. Waxenbaum, V. Reddy, and M. Varacallo, "Anatomy, Autonomic Nervous System," in *StatPearls*, Treasure Island (FL): StatPearls Publishing, 2023.
[3] G. Grassi, "Role of the sympathetic nervous system in human hypertension," *Journal of hypertension*, vol. 16, no. 12, pp. 1979–1987, 1998.
[4] M. Sinski, J. Lewandowski, P. Abramczyk, K. Narkiewicz, and Z. Gaciong, "Why study sympathetic nervous system," *J Physiol Pharmacol*, vol. 57, no. Suppl 11, pp. 79–92, 2006.
[5] S. Ewen, C. Ukena, D. Linz, R. E. Schmieder, M. Böhm, and F. Mahfoud, "The sympathetic nervous system in chronic kidney disease," *Current hypertension reports*, vol. 15, pp. 370–376, 2013.
[6] Y. Kong and K. H. Chon, "Electrodermal activity in pain assessment and its clinical applications," *Applied Physics Reviews*, vol. 11, no. 3, p. 031316, Aug. 2024, doi: 10.1063/5.0200395.
[7] N. Gullett, Z. Zajkowska, A. Walsh, R. Harper, and V. Mondelli, "Heart rate variability (HRV) as a way to understand associations between the autonomic nervous system (ANS) and affective states: A critical review of the literature," *International Journal of Psychophysiology*, 2023
[8] J. F. Thayer, S. S. Yamamoto, and J. F. Brosschot, "The relationship of autonomic imbalance, heart rate variability and cardiovascular disease risk factors," *International journal of cardiology*, vol. 141, no. 2, pp. 122–131, 2010.
[9] T. Kusayama *et al.*, "Simultaneous noninvasive recording of electrocardiogram and skin sympathetic nerve activity (neuECG)," *Nature protocols*, vol. 15, no. 5, pp. 1853–1877, 2020.
[10] M. Ogawa *et al.*, "Cryoablation of stellate ganglia and atrial arrhythmia in ambulatory dogs with pacing-induced heart failure," *Heart Rhythm*, vol. 6, no. 12, pp. 1772–1779, 2009.
[11] G. Feigin, S. Velasco Figueroa, M. F. Englesakis, R. D'Souza, Y. Hoydonckx, and A. Bhatia, "Stellate ganglion block for non-pain indications: a scoping review," *Pain Medicine*, vol. 24, no. 7, pp. 775–781, 2023.
[12] F. Baghestani, Y. Kong, W. D'Angelo, and K. H. Chon, "Analysis of Sympathetic Responses to Cognitive Stress and Pain through Skin Sympathetic Nerve Activity and Electrodermal Activity," *Computers in Biology and Medicine*, 2024.
[13] Y. Xing *et al.*, "An artifact-resistant feature SKNAER for quantifying the burst of skin sympathetic nerve activity signal," *Biosensors*, vol. 12, no. 5, p. 355, 2022.
[14] P. Kligfield *et al.*, "Recommendations for the Standardization and Interpretation of the Electrocardiogram: Part I: The Electrocardiogram and Its Technology: A Scientific Statement From the American Heart Association Electrocardiography and Arrhythmias Committee, Council on Clinical Cardiology; the American College of Cardiology Foundation; and the Heart Rhythm Society Endorsed by the International Society for Computerized Electrocardiology," *Circulation*, vol. 115, no. 10, pp. 1306–1324, Mar. 2007, doi: 10.1161/CIRCULATIONAHA.106.180200.
[15] J. S. Burma, A. P. Lapointe, A. Soroush, I. K. Oni, J. D. Smirl, and J. F. Dunn, "Insufficient sampling frequencies skew heart rate variability estimates: Implications for extracting heart rate metrics from neuroimaging and physiological data," *Journal of Biomedical Informatics*, vol. 123, p. 103934, Nov. 2021, doi: 10.1016/j.jbi.2021.103934.
[16] R. E. Gregg, S. H. Zhou, J. M. Lindauer, E. D. Helfenbein, and K. K. Giuliano, "What is inside the electrocardiograph?," *Journal of electrocardiology*, vol. 41, no. 1, pp. 8–14, 2008.
[17] F. Baghestani, M. P. S. Nejad, Y. Kong, and K. H. CHon, "Towards Continuous Skin Sympathetic Nerve Activity Monitoring: Removing Muscle Noise Artifacts," in *IEEE-EMBS International Conference on Body Sensor Networks (IEEE BSN)*, 2024.
[18] Y. Kong, F. Baghestani, W. D'Angelo, I.-P. Chen, and K. H. Chon, "A *Novel* Approach to Characterize Dynamics of ECG-Derived Skin Nerve Activity via Time-Varying Spectral Analysis," Nov. 13, 2024, *arXiv*: arXiv:2411.08308. doi: 10.48550/arXiv.2411.08308.
[19] H. Wang, K. Siu, K. Ju, and and K. H. Chon, "A High Resolution Approach to Estimating Time-Frequency Spectra and Their Amplitudes," *Ann Biomed Eng*, vol. 34, no. 2, pp. 326–338, Feb. 2006.
[20] J. N. Mandrekar, "Receiver Operating Characteristic Curve in Diagnostic Test Assessment," *Journal of Thoracic Oncology*, vol. 5, no. 9, pp. 1315–1316, Sep. 2010, doi: 10.1097/JTO.0b013e3181ec173d.
[21] T. K. Koo and M. Y. Li, "A Guideline of Selecting and Reporting Intraclass Correlation Coefficients for Reliability Research," *Journal of Chiropractic Medicine*, vol. 15, no. 2, pp. 155–163, Jun. 2016, doi: 10.1016/j.jcm.2016.02.012.